\documentclass[longauth]{aa}  

\usepackage{graphicx}
\usepackage{txfonts}
\usepackage{orcidlink}
\usepackage{placeins} 

\begin{document}

   \title{TOI-512: Super-Earth transiting a K-type star discovered by TESS and ESPRESSO\thanks{Based on Guaranteed Time Observations collected at the European Southern Observatory by the ESPRESSO Consortium under ESO programmes 106.21M2.007, 108.2254.002, 108.2254.005, and 108.2254.006.}}

   \subtitle{}

   \author{J.\,Rodrigues 
          \inst{\ref{IA},\ref{DFA},\ref{OFXB}}\thanks{\email{jose.rodrigues@astro.up.pt}}\orcidlink{0000-0001-5164-3602}
          \and S.\,C.\,C.\,Barros\inst{\ref{IA},\ref{DFA}}\orcidlink{0000-0003-2434-3625}
          \and N.\,C.\,Santos \inst{\ref{IA},\ref{DFA}}\orcidlink{0000-0003-4422-2919}
          \and J.\,Davoult \inst{\ref{CSH},\ref{PhBern}}\orcidlink{0000-0002-6177-2085}
          \and M.\,Attia \inst{\ref{ObsGe}}\orcidlink{0000-0002-7971-7439}
          \and A.\,Castro-Gonz\'{a}lez \inst{\ref{INAF-Torino}}\orcidlink{0000-0001-7439-3618}
          \and S.\,G.\,Sousa \inst{\ref{IA}}\orcidlink{0000-0001-9047-2965}
          \and O.\,D.\,S.\,Demangeon \inst{\ref{IA},\ref{DFA}}\orcidlink{0000-0001-7918-0355}
          \and M.\,J.\,Hobson \inst{\ref{ObsGe}}\orcidlink{0000-0002-5945-7975}
          \and D.\,Bossini \inst{\ref{IA},\ref{DFA}} 
          \and C.\,Ziegler \inst{\ref{AustinState}}
          \and J.\,P.\,Faria \inst{\ref{ObsGe}}\orcidlink{0000-0002-6728-244X}
          \and V.\,Adibekyan \inst{\ref{IA}}\orcidlink{0000-0002-0601-6199}
          \and C.\,Lovis \inst{\ref{ObsGe}}\orcidlink{0000-0001-7120-5837}
          \and B.\,Lavie \inst{\ref{ObsGe}}\orcidlink{0000-0001-8884-9276}
          \and M.\,Damasso \inst{\ref{INAF-Torino}}\orcidlink{0000-0001-9984-4278}
          \and A.\,M.\,Silva \inst{\ref{IA},\ref{DFA}}\orcidlink{0000-0003-4920-738X}
          \and A.\,Suárez Mascareño \inst{\ref{IAC}}\orcidlink{0000-0002-3814-5323}
          \and F.\,Pepe \inst{\ref{ObsGe}}
          \and F.\,Bouchy \inst{\ref{ObsGe}}\orcidlink{0000-0002-7613-393X}
          \and Y.\,Alibert \inst{\ref{CSH},\ref{PhBern}}\orcidlink{0000-0002-4644-8818}
          \and J.\,I.\,Gonz\'alez Hern\'andez \inst{\ref{IAC},\ref{ULaguna}}\orcidlink{0000-0002-0264-7356}
          \and A.\,Sozzetti \inst{\ref{INAF-Torino}}\orcidlink{0000-0002-7504-365X}
          \and C.\,Allende Prieto \inst{\ref{IAC}}\orcidlink{0000-0002-0084-572X}
          \and S.\,Cristiani \inst{\ref{INAF-Trieste}, \ref{IFPU-Trieste}}\orcidlink{0000-0002-2115-5234}
          \and E.\,Palle \inst{\ref{IAC}, \ref{ULaguna}}\orcidlink{0000-0003-0987-1593}
          \and V.\,D'Odorico \inst{\ref{INAF-Trieste}}\orcidlink{0000-0003-3693-3091}
          \and D.\,Ehrenreich \inst{\ref{ObsGe}}\orcidlink{0000-0001-9704-5405}
          \and P.\,Figueira \inst{\ref{ObsGe},\ref{IA}, \ref{ESO-Chile}}\orcidlink{0000-0001-8504-283X}
          \and{K.\,G.\,Stassun \inst{\ref{Vanderbilt}}}\orcidlink{0000-0002-3481-9052}
          \and R.\,Génova Santos \inst{\ref{IAC},\ref{ULaguna}}\orcidlink{0000-0001-5479-0034}
          \and G.\,Lo Curto \inst{\ref{ESO-Garching}}\orcidlink{0000-0002-1158-9354}
          \and C.\,J.\,A.\,P.\,Martins \inst{\ref{IA},\ref{CAUP}}\orcidlink{0000-0002-4886-9261}
          \and A.\,Mehner \inst{\ref{ESO-Chile}}\orcidlink{0000-0002-9564-3302}
          \and G.\,Micela \inst{\ref{INAF-Palermo}}\orcidlink{0000-0002-9900-4751}
          \and P.\,Molaro \inst{\ref{INAF-Trieste}}\orcidlink{0000-0002-0571-4163}
          \and N.\,J.\,Nunes \inst{\ref{IA-Lisboa}}\orcidlink{0000-0002-3837-6914}
          \and E.\,Poretti \inst{\ref{INAF-TNG}}\orcidlink{0000-0003-1200-0473}
          \and R.\,Rebolo \inst{\ref{IAC}}\orcidlink{0000-0003-3767-7085}
          \and S.\,Udry \inst{\ref{ObsGe}}\orcidlink{0000-0001-7576-6236}
          \and M.\,R.\,Zapatero Osorio\inst{\ref{CAB-Madrid}}\orcidlink{0000-0001-5664-2852}
          }

   \institute{Instituto de Astrofísica e Ciencias do Espaço, Universidade do
            Porto, CAUP, Rua das Estrelas, PT4150-762 Porto, Portugal\label{IA}    
         \and
             Departamento de Fisica e Astronomia, Faculdade de Ciencias, Universidade do Porto, Rua Campo Alegre, 4169-007 Porto, Portugal\label{DFA}
        \and
            Observatoire François-Xavier Bagnoud -- OFXB, 3961 Saint-Luc, Switzerland\label{OFXB}
        \and 
            Center for Space and Habitability, University of Bern, Gesellschaftsstrasse\,6, CH-3012 Bern, Switzerland\label{CSH}
        \and
            Physics Institute of University of Bern, Gesellschaftsstrasse\,6, CH-3012 Bern, Switzerland\label{PhBern}
        \and
            Observatoire Astronomique de l'Université de Genève, Chemin Pegasi 51b, 1290 Versoix, Switzerland\label{ObsGe}
        \and
            INAF -- Osservatorio Astrofisico di Torino, Via Osservatorio\,20, I-10025 Pino Torinese, Italy\label{INAF-Torino}
        \and
            INAF -- Osservatorio Astrofisico di Trieste, via G. B. Tiepolo 11, 34143, Trieste, Italy\label{INAF-Trieste}     
        \and 
            IFPU--Institute for Fundamental Physics of the Universe, via Beirut 2, I-34151 Trieste, Italy\label{IFPU-Trieste}
        \and
            Department of Physics, Engineering and Astronomy, Stephen F. Austin State University, 1936 North St, Nacogdoches, TX 75962, USA\label{AustinState}
        \and
            Instituto de Astrof{\'\i}sica de Canarias, E-38205 La Laguna, Tenerife, Spain\label{IAC}
        \and
            Universidad de La Laguna, Dept. Astrof{\'\i}sica, E-38206 La Laguna, Tenerife, Spain\label{ULaguna}
        \and
            European Southern Observatory, Karl-Schwarzschild-Strasse 2, 85748, Garching b. Munchen, Germany\label{ESO-Garching}
        \and
            European Southern Observatory, Alonso de Cordova 3107, Vitacura, Region Metropolitana, Chile\label{ESO-Chile}
        \and 
            Centro de Astrofísica da Universidade do Porto, Rua das Estrelas, 4150-762 Porto, Portugal\label{CAUP}
        \and
            Department of Physics and Astronomy, Vanderbilt University, Nashville, TN 37235, USA\label{Vanderbilt}
        \and
            INAF – Osservatorio Astronomico di Palermo, Piazza del Parlamento 1, 90134 Palermo, Italy\label{INAF-Palermo}
        \and
            Fundación Galileo Galilei-INAF, Rambla José Ana Fernandez Pérez 7, 38712 Breña Baja, TF, Spain\label{INAF-TNG}
        \and
            Instituto de Astrofísica e Ciencias do Espaco, Faculdade de Ciencias da Universidade de Lisboa, Edifício C8, Campo Grande, 1749-016 Lisbon, Portugal\label{IA-Lisboa}
        \and
            Centro de Astrobiología, CSIC-INTA, Camino Bajo del Castillo s/n, 28602 Villanueva de la Cañada, Madrid, Spain\label{CAB-Madrid}
        }

   \date{Received 5 November 2024 / Accepted 10 February 2025}

 
  \abstract
   {One of the goals of the ESPRESSO guaranteed time observations (GTOs) at the ESO 8.2m telescope is to follow up on candidate planets from transit surveys such as the TESS mission. High-precision radial velocities are required to characterize small exoplanets.} 
   {We intend to confirm the existence of a transiting super-Earth around the bright (V=9.74) K0-type star TOI-512 (TIC 119292328) and provide a characterization.}
   {Combining photometric data from TESS and 37 high-resolution spectroscopic observations from ESPRESSO in a joint Markov chain Monte Carlo analysis, we determined the planetary parameters of TOI-512b and characterized its internal structure.}
   {We find that TOI-512b is a super-Earth, with a radius of $1.54 \pm 0.10$ R$_\oplus$ and mass of $3.57_{-0.55}^{+0.53}$~M$_\oplus$, on a $7.19_{-6.1\cdot 10^{-5}}^{+7\cdot 10^{-5}}$ day orbit. This corresponds to a bulk density of $5.62_{-1.28}^{+1.59}$ g cm$^{-3}$. 
    Our interior structure analysis presents a small inner core representing $0.13^{+0.13}_{-0.11}$ of the solid mass fraction for the planet, surrounded by a mantle with a mass fraction of $0.69^{+0.20}_{-0.22}$, and an upper limit of the water layer of $0.16$. The gas mass below $10^{-8.93}$ indicates a very small amount of gas on the planet. We find no evidence of the second candidate found by the TESS pipeline, TOI-512.02, neither in TESS photometry, nor in the ESPRESSO radial velocities. The low stellar activity makes it an interesting transmission spectroscopy candidate for  future-generation instruments.}
   {}


   \maketitle
%


\section{Introduction}
\label{sec:intro}

    Transiting exoplanets are especially valuable as their geometry enables us to derive accurate planetary properties, including their mass and radius. 
    Following the great success of radial velocity surveys (e.g., \citealt{Mayor2011_Harps}) and the advent of space-based photometric surveys such as \textit{Kepler} \citep{Borucki_2010_Kepler}, K2 \citep{Howell_2014_K2}, and the Transiting Exoplanet Survey Satellite \citep[TESS;][]{Ricker_2015_TESS}, the number of systems hosting planetary candidates has swiftly risen into the thousands over a short amount of time. 
    Transit data, however, typically only permit the derivation of the radius of the planet. Follow-up radial velocity measurements are usually needed to derive masses. Furthermore, transit surveys are subject to a variety of phenomena that often produce false positive detections \citep{Fressin_2013_FalsePositives, FAP2021MNRAS.508..195D, FAP2022AJ....163..244C, FAP2022MNRAS.509.1075C} and require dedicated follow-up observations to establish their true nature.
    
    The  high-resolution spectrograph ESPRESSO \citep{Pepe_espresso_vlt_2021} on board Very Large Telescope (VLT) has been a stepping stone towards precise determination of planetary masses. The instrument has shown its ability to achieve stable radial velocity (RV) measurements, with a precision of a few tens of cm~s$^{-1}$ over the course of months \citep{Demangeon2021_TOI-175, Faria2022_proxima}.
    One of the goals of the ESPRESSO Guaranteed Time Observations (GTO) consortium is the precise characterization of transiting planetary systems discovered by K2 and TESS. Combining high-precision radius measurements with ESPRESSO RV observations allows us to precisely constrain the fundamental physical properties of small exoplanets and to infer their bulk densities, offering insights into their composition, as demonstrated by recent work of the GTO team \citep{damasso_pi_mensae_2020, sozzetti2021_TOI-130, Demangeon2021_TOI-175, Lavie2023_TOI-455, Barros2022_TOI-174, Damasso2023-TOI-469, Castro-Gonzalez2023-TOI-244, Hobson2024_WG3}. 

    The California-\textit{Kepler} Survey exposed a bi-modality of the radius distribution of transiting planets, showing a lack of them in the 1.5-2$R_\oplus$ range \citep{Fulton2017, Fulton2018}.
    This feature of the radius distribution could be explained by XUV photoevaporation \citep{Lammer2003photoevaporation, Lopez2013photoevap}, where the flux from the host star heats up the higher atmosphere of the exoplanet to temperatures around $\sim10^4 K$, causing an outflow of H/He-rich envelopes and thereby transforming sub-Neptunes into naked rocky cores. As a result of irradiation, we expect inner planets to become denser and smaller \citep{HoweBurrows2015}. 
    Another possible mechanism that could explain this radius valley is core-powered mass-loss \citep{Ginzburg2016corepowered,GuptaSchlichting2020Corepowered,Rogers2021photoVScore-pow}, where thermal energy from the planet formation is slowly released from the core into the atmosphere, causing it to radiate away.
    Changes in dust properties beyond the water-ice line could also play a significant role in the origin of the radius valley \citep{Venturini2020}.
    
    In 2020, the TESS mission announced the detection of the TOI-512 system. ExoFOP\footnote{\url{https://exofop.ipac.caltech.edu/tess/}} reported two candidates; TOI-512.01 and TOI-512.02, with radii of $1.53~R_\oplus$ and $1.74~R_\oplus$, and insolations of 112 and 57 that of the Earth, respectively. Those parameters would place them right in the aforementioned radius gap, with a stellar incident flux where both gas-rich sub-Neptunes and gas-poor rocky planets appear to coexist. Therefore, obtaining precise RV observations to characterize the nature of the candidates was deemed necessary. This system was thus observed in the context of the ESPRESSO GTO.
    
    This paper describes the discovery and characterization of a new transiting super-Earth with TESS and ESPRESSO. We report no detection of TOI-512.02 with photometry, nor in the radial velocities. Therefore, we have ruled out its existence. The observations of this target and its host star are summarized in Sect. \ref{sec:obs}. The derivation of stellar parameters, initial photometric, and spectroscopic analyses, along with the joint modeling are described in Sect. \ref{sec:data_analysis}. The internal structure is analyzed in Sect. \ref{sec:internal_structure}, while the prospects of spectroscopic follow-up observations are presented in Sect. \ref{sec:spectroFollow-up}. The conclusions are given in Sect. \ref{sec:conclusion}.
    

\section{Observations}
\label{sec:obs}
\subsection{TESS photometry}

    TOI-512 (TIC 119292328, CD--38 2650, Gaia DR3 5574111674620962688) was observed by TESS during sector 6 (2018 December 11 -- 2019 January 7), sector 7 (2019 January 08 -- 2019 February 01), and sector 33 (2020 December 17 -- 2021 January 13) on camera 2, in the 2-min cadence mode ($t_{exp} =$ 2~min). The observations were processed using the Science Processing Operations Center (SPOC) pipeline \citep{SPOC2016}. We retrieved the data through the \texttt{lightkurve} Python package \citep{Lightkurve2018}. For our analysis, we used the Presearch Data Conditioning Simple Aperture Photometry (PDCSAP) flux \citep{PDCSAP2012, PDCSAP2014, PDCSAP2012II}, which has common instrumental trends removed. The light curve was constructed and analyzed using the \texttt{FULMAR}\footnote{\url{https://fulmar-astro.readthedocs.io/en/latest/}} Python package \citep{FULMAR}.
    
    The PDCSAP light curve consists of 48492 2-min exposures with a median uncertainty of 596~parts per million (ppm). We removed observations more than five times the standard deviation above the mean of the observed flux. To avoid removing transits, a similar approach was used for observations more than five times the standard deviation under the mean flux -- with the difference that each observation, $o_i$, whose two neighbors $\{o_{i-1}, o_{i+1}\}$ were considered outliers as well were considered to be of a plausibly physical origin; therefore, they were not removed from the light curve.
    The individual sector light curves were normalized before being combined to produce the final light curve. Observations flagged by the pipeline as low-quality were removed. 
    
    Using \texttt{FULMAR}, we started the analysis of the TESS light curve by computing a \texttt{Transit Least-Squares} (TLS) periodogram \citep{HippkeTLS2019}. TOI-512.01 has been recovered with a high (32.04) signal detection efficiency (SDE). 

    The candidate has a reported orbital period of $7.1885 \pm 0.00205 $~days and a transit depth of $277 \pm 31$~ppm, corresponding to a radius of $1.58 \pm 0.09 R_{\oplus}$. A second candidate was reported in 2021 with an orbital period of $20.27494 \pm 0.00045 $~days and a transit depth of $410 \pm 1.8$~ppm, corresponding to a radius of $1.74 \pm 0.16 R_{\oplus}$. We report no detection of this second candidate, neither with photometry nor in the radial velocities.

    We masked data spanning 1.2 times the duration of each transit from the light curve before conducting a second TLS search, which did not recover any significant signal. Searching by sector combination (6-7, 6-33 and 7-33) did not lead to a different result. A visual analysis of the folded SPOC (shown in Fig. \ref{fig:transitcheck}) and the QLP light curves was not convincing either. An independent search was conducted using the method described in \citet{Barros2016K2}, which also lead to no detection. 
    
\begin{figure}
        \centering
        \includegraphics[width=\linewidth]{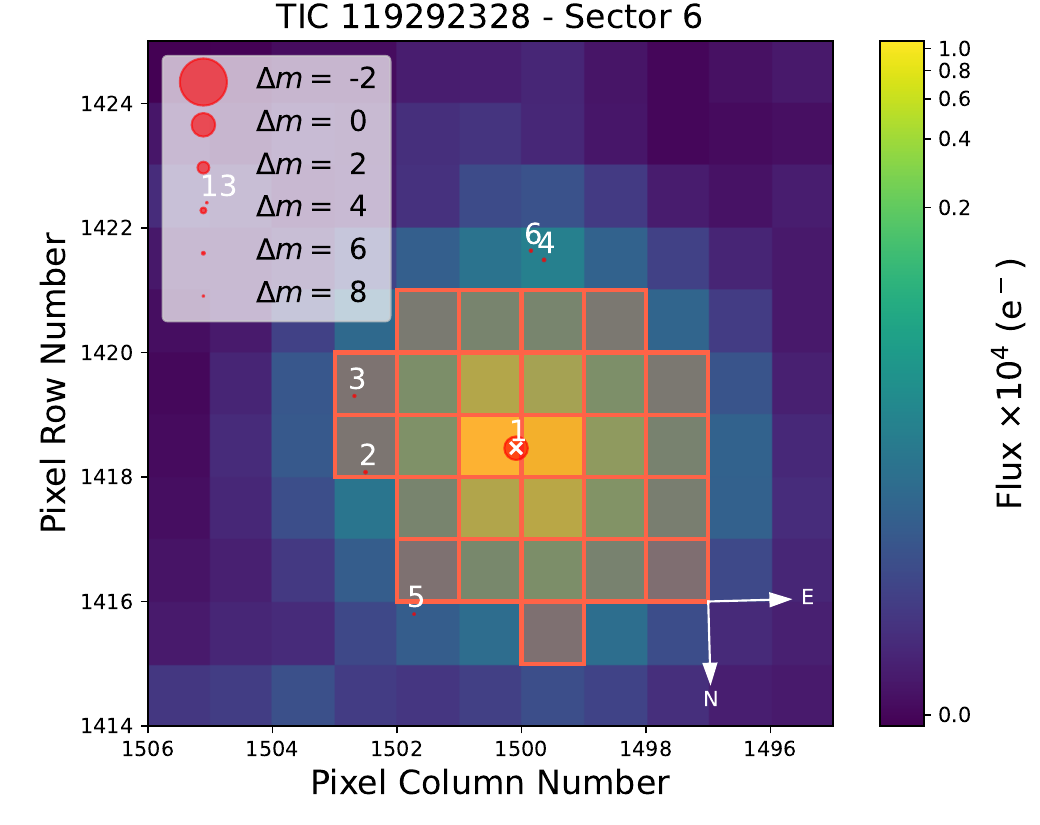}
        \caption{TESS target pixel files for sector 6 \textsc{tpfplotter} (\citealt{Aller_tpfplotter2020}; the code is publicly available at www.github.com/jlillo/tpfplotter). Orange squares identify the aperture masks used to extract the light curve. Sources cross-matched with the {\it Gaia} DR3 catalog are indicated by red dots, whose sizes are scaled with their relative magnitude, compared to that of TOI-512. The pixel scale is 21$\arcsec$ pixel$^{-1}$}
        \label{fig:tpfplotter}
    \end{figure}

    Figure \ref{fig:tpfplotter} shows the TESS target pixel files (TPF) centered on TOI-512, with sources cross-matched with the {\it Gaia} DR3 catalog overplotted. We used the TESS-cont algorithm \citep{Castro-Gonzalez2024-TESS-cont} to quantify the contamination from nearby \textit{Gaia} sources inside the photometric aperture. We find that 99.4\% of the flux comes from TOI-512, which ensures negligible contamination.

\subsection{Imaging observations}
    \label{sec:imaging}

    High-angular resolution imaging is needed to search for nearby sources that can contaminate the TESS photometry, resulting in an underestimated planetary radius, or be the source of astrophysical false positives, such as background eclipsing binaries. We searched for stellar companions to TOI-512 with speckle imaging on the 4.1-m Southern Astrophysical Research (SOAR) telescope \citep{tokovinin2018SOAR}. TOI-512 was observed on 18 May 2019 in the Cousins I-band, a similar visible bandpass to TESS. This observation was sensitive to a 5.2-magnitude fainter star at an angular distance of 1 arcsec from the target. More details of the observations within the SOAR TESS survey are available in \citet{ziegler2020_SOAR_TESS}.
    
\begin{figure}
        \resizebox{\hsize}{!}{\includegraphics{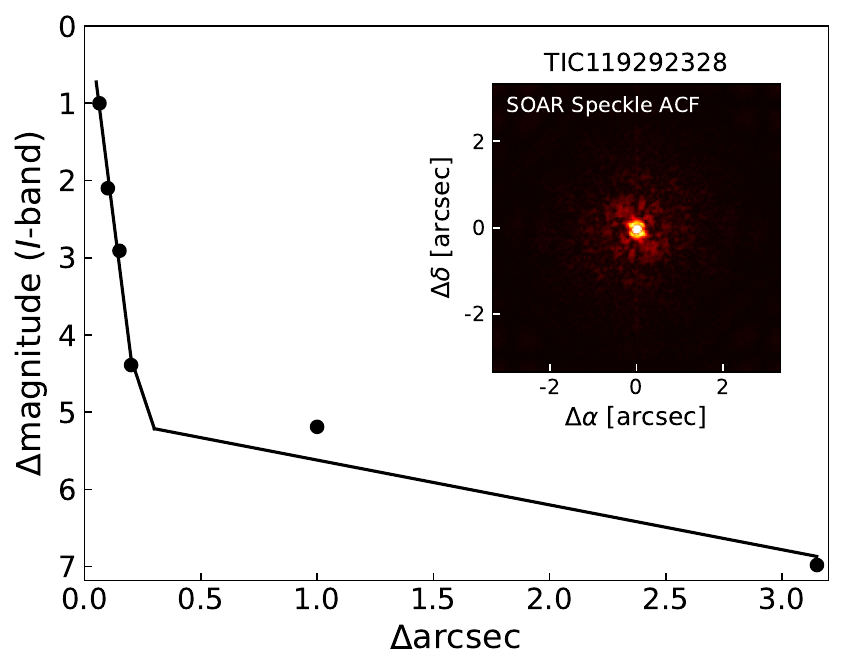}}
        \caption{Speckle imaging of TOI-512 taken with the SOAR telescope on 18 May 2019. No nearby stars were detected within 3" of the target.}
        \label{fig:SOAR}
    \end{figure}

    No nearby stars were detected within 3$\arcsec$ of TOI-512 in the SOAR observations. The 5$\sigma$ detection sensitivity and speckle auto-correlation functions (ACF) from the observations are shown in Fig. \ref{fig:SOAR}.
    
\subsection{ESPRESSO observations}
    \label{sec:obs_ESPRESSO}

    We collected 37 high-resolution spectra of TOI-512 (K0V, V=9.74) between 2021 September 2 and 2022 April 28 using the ESPRESSO spectrograph (\citealt{Pepe_espresso_vlt_2021}),  based at the Paranal Observatory. The observations were obtained as part of the ESPRESSO Guaranteed Time Observations subprogram aimed at measuring precise masses of transiting planet candidates uncovered by the TESS and K2 missions (Programs: 106.21M2.007, 108.2254.002, 108.2254.005, and 108.2254.006, PI: F.Pepe). The observations were carried out with Fabry-Pérot (FP) simultaneous calibration. We used single UT high-resolution mode (HR21, fast-readout) for these observations, covering wavelengths from 380 nm to 788 nm with a spectral resolution of $R = 140\, 000$. The S/N is 60 per resolution element at 650nm.
    
    The RVs were extracted using the ESPRESSO data-reduction software (DRS) version 3.0.0. It computes the CCF of the sky-subtracted spectra with a stellar line mask \citep{Baranne_ELODIE_1996, Pepe2000} to calculate the RVs. The mask was optimized for G9 stars in our case. The DRS then fits the CCF with an inverted Gaussian profile. The parameters of the profile are the center of the profile, which gives the measurement of the RV; its FWHM; the amplitude (contrast of the CCF); and the bissector span. The uncertainties on the RVs are computed following the technique described in \citet{Bouchy2001}. In addition to RV, the DRS computes other activity indicators: the BIS \citep{Queloz_BIS_2001}, the depth of the $H_\alpha$ line \citep{GomesDaSilva2011}, the depth of the sodium doublet NaD \citep{Diaz_NaD_2007}, the S-Index \citep{Noyes_S-Index_1984, Lovis_S-index_2011} and the $\log R'_{HK}$ \citep{BoisseLogRHK_2010} The median uncertainty of the RV measurements is 0.35~ms$^{-1}$. 
    
\begin{figure*}
        \centering
        \includegraphics[width=\textwidth]{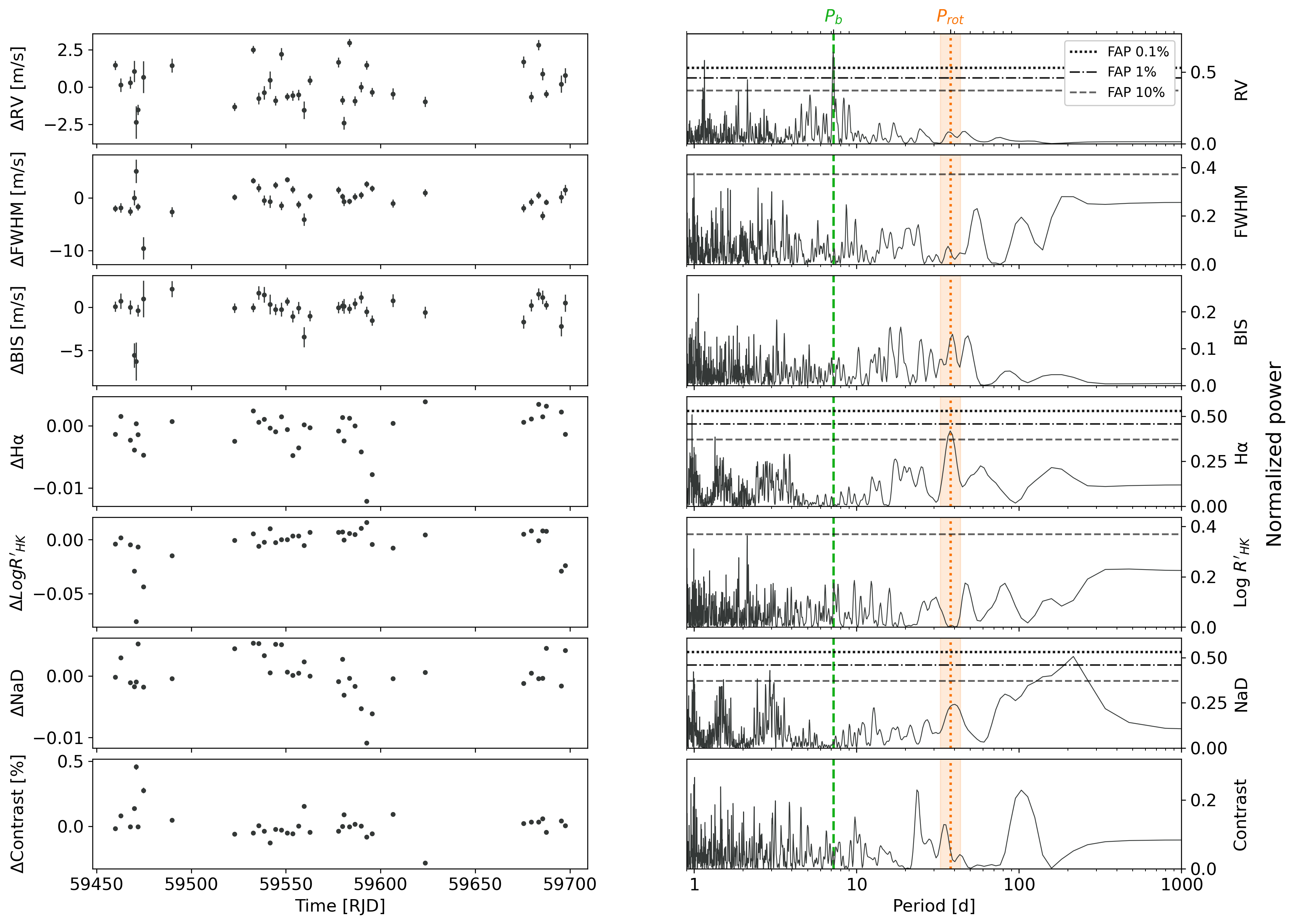}
        \caption{\textit{Left panel:} Time series of the RV observations and spectroscopic activity indicators presented in Sec.\ref{sec:obs_ESPRESSO}. \textit{Right panel:} GLS periodograms of the RV and activity indicators' timeseries. Horizontal lines represent the 0.1\%, 1\%, and 10\% false alarm probability (FAP) derived following \citet{Baluev_2008}. The green dashed vertical line represents the period of TOI-512b. The orange dotted vertical line represents the stellar rotational period from \citet{Suarez-Mascareno_ha_2015}}
        \label{fig:timeseries_periodograms}
    \end{figure*}

    We show the time series of the RV and activity indicators (FWHM, BIS, $H_\alpha$, $\log R'_{KH}$, NaD, and Contrast) and their respective generalized Lomb-Scargle (GLS) periodograms \citep{Zechmeister_GLS_2009} in Fig.~\ref{fig:timeseries_periodograms}. The candidate signal TOI-512.01 is detected in the RVs, with a false alarm probability of $10^{-4.1}$; thus, hereinafter, we refer to it as TOI-512 b. We see no significant peak at the period of the planet in any of the stellar activity proxies. The candidate signal TOI-512.02 is not detected in the RVs either.

    The GLS periodogram of the $H_\alpha$ timeseries shows the most prominent peak at $0.97$~days with a FAP of $0.21\%$. This peak is a one-day alias of the $37.72$~days, which shows a $3\%$ FAP. This period is consistent with the expected rotational period of the star computed using Eq. (9) of \citet{Suarez-Mascareno_ha_2015}, together with our $\log R'_{HK}$ measurement of $37.79^{+5.75}_{-4.99}$ days. We therefore conclude that its origin lies in stellar rotation.
    
\section{Data analysis and system parameters}
\label{sec:data_analysis}
\subsection{Stellar parameters}
    \label{sec:stellar_params}
   
    We combined 37 ESPRESSO spectra to derive the stellar atmospheric parameters following the methodology described in \citet{Sousa_SweetCat2_2021,Sousa2014, Santos_SweetCat_2013}. The \texttt{ARES}\footnote{\url{http://www.astro.up.pt/\~sousasag/ares}} code \citep{Sousa_ARES_2007, Sousa_ARES_v2_2015} was used to measure the equivalent widths (EWs) of iron lines in the combined ESPRESSO spectrum. The radiative transfer code \texttt{MOOG} \citep{SnedenPhD1973} was then used on a grid of Kurucz model atmospheres \citep{Kurucz1993} to estimate the abundances. Assuming ionization and excitation equilibrium leads to the best set of spectroscopic parameters. The stellar abundances of Mg and Si were determined using the classical curve-of-growth analysis, closely following the methods described in \citet{Adibekyan2012, Adibekyan_2015}. We used the {\it Gaia} DR3 parallax \citep{Gaia_DR3_2021} to estimate the trigonometric surface gravity at $4.44 \pm 0.03$~dex. 
    
    The stellar mass $M_\star= 0.74 \pm 0.03~M_\odot$,  the stellar radius $R_\star = 0.89 \pm 0.03~R_\odot$, and stellar age $8.235 \pm 4.386$~Gyr were derived using the Bayesian tool \texttt{PARAM}\footnote{\url{http://stev.oapd.inaf.it/cgi-bin/param}} \citep{DaSilva_2006, Rodrigues14PARAM,Rodrigues17PARAM}, which matches spectroscopic parameters $T_{eff}$ and $[Fe/H]$, and the stellar luminosity to a well-sampled grid of stellar evolutionary tracks and isochrones from the \texttt{PARSEC}\footnote{\url{http://stev.oapd.inaf.it/cgi-bin/cmd}} code \citep{Bressan2012_isochrones}. We assumed a stellar age between 1~Myr and 14~Gyr, a constant star formation rate and an initial mass function from \citet{Chabrier2001}. To determine the stellar luminosity, we used the {\it Gaia} parallax as a proxy for distance, the 2MASS $K_s$ magnitude \citep{2MASS_2006}, and the bolometric correction deduced by \texttt{YBC}\footnote{\url{http://stev.oapd.inaf.it/YBC/}} \citep{YBC_Chen_19}. \texttt{YBC} interpolates pre-computed bolometric correction tables with the observed effective temperature, surface gravity, metallicity and extinction using a combination of ATLAS9 \citep{CastelliKurucz_ATLAS9_03} and Phoenix \citep{Allard_Phoenix_12} spectral libraries. The extinction was determined using \texttt{STILISM}\footnote{\url{https://stilism.obspm.fr/}} \citep{Lallament_STILISM_14, Capitanio_STILISM_17}. However, it was determined to be negligible due to the proximity of the star.
    
    For comparison, we used the activity-age relation of \citet{Mamajek_activity_age_2008} to determine the stellar age from spectral activity indicators, namely, $\log R'_{HK}$. The result of $8.44_{-1.84}^{+1.56}$~Gyr is consistent with the age derived from isochrone fitting, indicating an old star (albeit  of solar metallicity).
    We assessed the spectral type of TOI-512 to be K0V from its metallicity, position in the Hertzsprung--Russell diagram with evolutionary tracks and comparison with reference stars \citep{Pecaut_k0_2013}. The stellar parameters are given in Table \ref{tab:spectroparams}.
    
    \begin{table}[h]
    \caption{Stellar parameters of TOI-512}
    \begin{tabular}{l c c}
        \hline
        Parameter & Value & Ref\\
        \hline
        Astrometric properties: & & \\
        RA (hh:mm:ss.ssss) & 06:26:35.20 & 1 \\
        DEC (hh:mm:ss.ssss) & -38:36:24.27 & 1 \\
        Parallax (mas) & $14.8818 \pm 0.0098$ & 1 \\
        Distance (pc) &  $67.196 \pm 0.044$ & 1 \\
        $\langle RV \rangle_\textrm{ESPRESSO}$ (m s$^{-1}$) & $68076.96 \pm 0.14$ & 0 \\\\

        Magnitudes: & & \\
        TESS & $8.9767 \pm 0.006$ & 3 \\
        B & $ 10.51 \pm 0.04$ & 2 \\
        V & $ 9.74 \pm 0.03$ & 2 \\
        G & $ 9.4877 \pm 0.0027$ & 1 \\
        J & $ 8.260 \pm 0.020$ & 4 \\
        H & $ 7.937 \pm 0.057$ & 4 \\
        K & $ 7.797 \pm 0.021$ & 4 \\\\
        
        Physical properties: & & \\
        $M_\star$ (M$_\odot$) & $0.74 \pm 0.03$ & 0 \\
        $R_\star$ (R$_\odot$) & $0.89 \pm 0.03$ & 0 \\
        $L_{\text{bol,}\star}$ & $0.5534 \pm 0.0068$ & 0 \\
        $T_{eff}$ (K) & $5277 \pm 67$ & 0 \\
        $\log g$ (g cm$^{-2}$)& $4.35 \pm 0.12$ & 0 \\
        $\log g_{\text{Gaia}}$ (g cm$^{-2}$)& $4.44 \pm 0.03$ & 1 \\
        Microturbulence (ms$^{-1}$) & $0.674 \pm 0.062$ & 0 \\
        $[Fe/H]$ (dex) & $ -0.11 \pm 0.05$ & 0 \\
        $[Mg/H]$ (dex) & $-0.02 \pm 0.04$ & 0 \\
        $[Si/H]$ (dex) & $-0.06 \pm 0.04$ & 0 \\
        $\log R'_{HK}$ (dex) & $-5.1 \pm 0.01$ & 0 \\
        $v \sin i$ (km s$^{-1}$)& $1.63 \pm 0.86$ & 0 \\ 
        Age (Gyr) & $8.235 \pm 4.386$ & 0 \\

        \hline
        
    \end{tabular}
    \tablefoot{We indicate the origin of the data with: 0 - This work, 1 - \citep{Gaia_DR3_2021}, 2 - \citep{Tycho-2Hipparcos}, 3 - \citep{TIC2019}, 4 - \citep{2MASS2003}.}
    \label{tab:spectroparams}
    \end{table}

\subsection{TESS photometry analysis}
\label{sec:photo_analysis}
    We analyzed the phase-folded light curve at the period corresponding to the highest peak of the TLS periodogram to assess its robustness. We then used \texttt{starry} \citep{starry2019} through the \texttt{exoplanet} package \citep{exoplanet:joss} to model the light curve and adopted a quadratic limb darkening law \citep{exoplanet:kipping13, exoplanet:agol20}. 
    We used a subsample of data centered around the mid-transit times, spanning 0.25 days before and after each transit, and we considered a circular orbit. 
    
    We used the period, $T_0$, and duration resulting from this fit to mask the data within three transit durations, centered around each transit. All transits were normalized by a linear trend computed from the out-of-transit data for our joint (RV+LC) final analysis.

\subsection{ESPRESSO Radial velocity analysis}
\label{sec:RV_analysis}
    We conducted a first independent analysis of the ESPRESSO data presented in Sect. \ref{sec:obs_ESPRESSO}. We computed the GLS periodogram of the RV, which shows a strong peak at $7.19$~days, with an associated FAP level of $10^{-4.1}$. A second peak with FAP < $1\%$ is visible at $1.16$~days, which is a one-day alias of the $7.19$~day signal.
    
\begin{figure*}
        \centering
        \includegraphics[width=\textwidth]{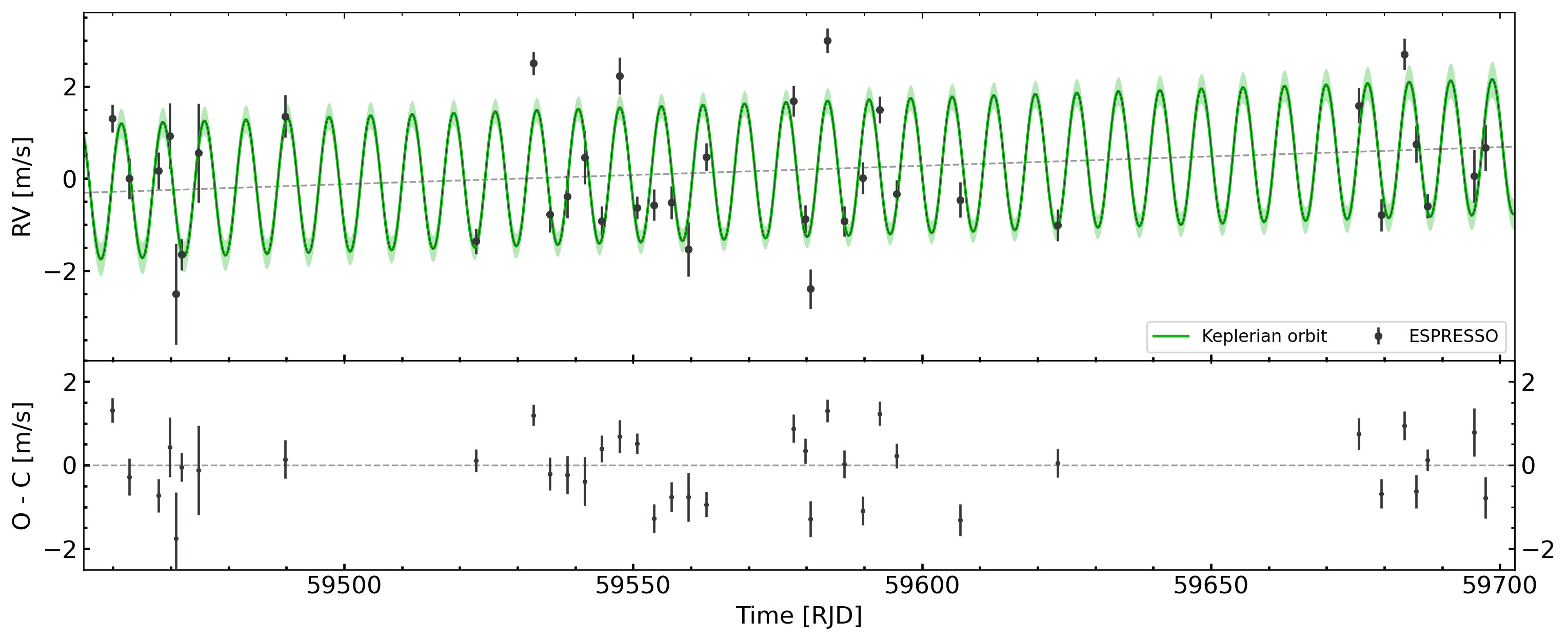}
        \caption{\textit{Top panel:} Time series of ESPRESSO RV observations of TOI-512. The green solid line and overlay display the best-fit RV component of the joint model presented in this study (cf. Sect. \ref{sec:joint_analysis}) with its relative 1-$\sigma$ uncertainty. \textit{Lower panel:} RV residuals.}
        \label{fig:timeseries_RV}
    \end{figure*}

\begin{figure}
        \resizebox{\hsize}{!}{\includegraphics{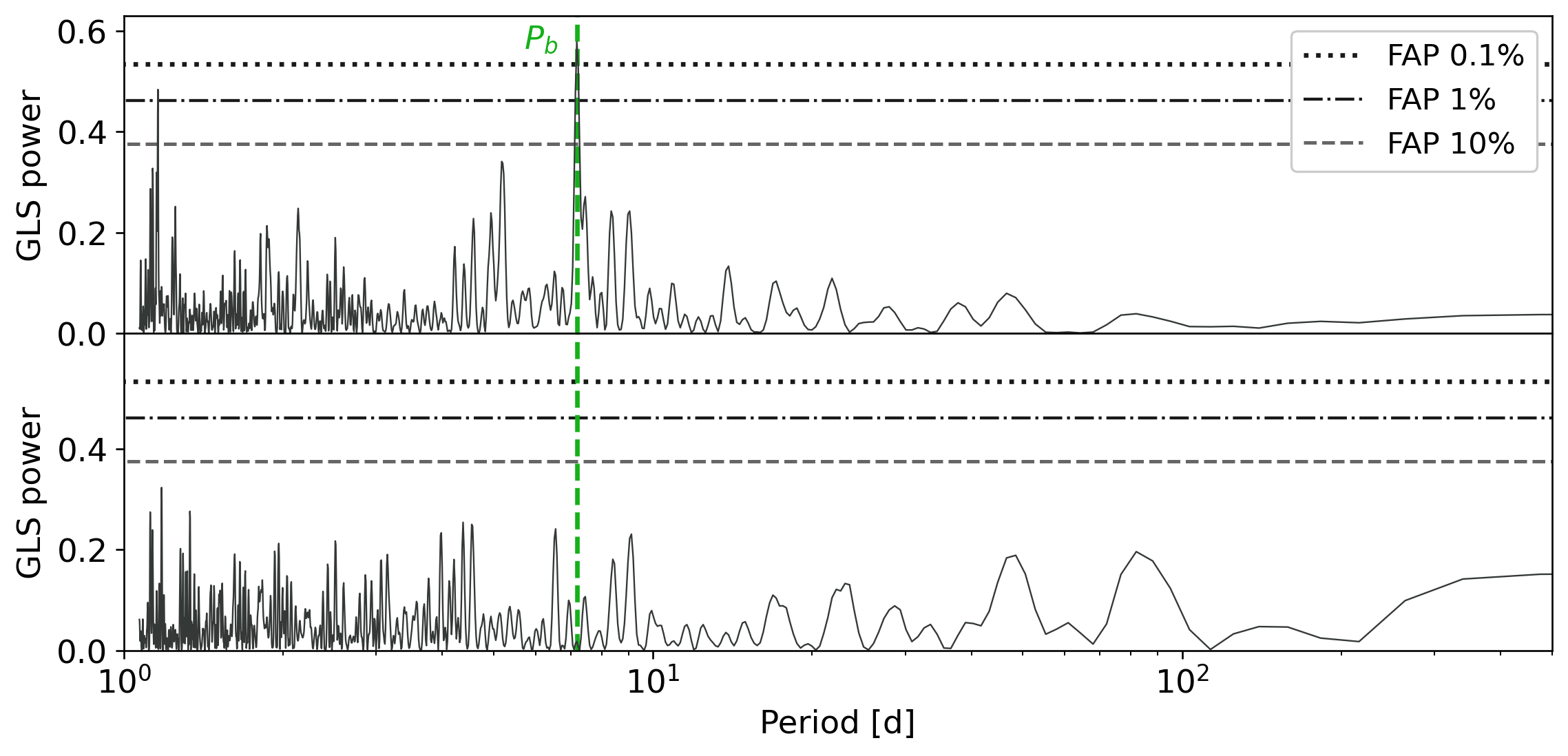}}
        \caption{\textit{Top panel:} GLS periodogram of the RV time series. \textit{Bottom panel panel:} GLS periodogram of the RV residuals. The green dashed vertical line represents the period of TOI-512b. Horizontal lines represent the 0.1$\%$, 1$\%,$ and 10$\%$ FAP.}
        \label{fig:RV_GLS_res}
    \end{figure}

    Figure~\ref{fig:timeseries_RV} shows the best-fit Keplerian model and its subtraction from the RV data. The root-mean-square (RMS) of the residuals is $0.81$~ms$^{-1}$. The GLS periodogram of the residuals (fig.~\ref{fig:RV_GLS_res}) shows no significant signal. There is no sign of another planet at the reported $20.27$~days in the RVs. There is a small ($1.24 \pm 0.34$ m s$^{-1}$ yr$^{-1}$) increasing trend in the RV time series. We could not attribute it to any known astrophysical source. As presented in Sect. \ref{sec:imaging}, adaptive optics (AO) imaging shows no evidence for a close stellar companion and the {\it Gaia} renormalized unit-wide error is below 1.4 (0.88), which would indicate a non-binary system. Our best guess for the origin of this signal is a long-term non-stellar companion.

\subsection{Radial velocity and light curve modeling}
\label{sec:joint_analysis}
    For the joint analysis of RV and photometric data presented in Sect. \ref{sec:obs}, we used the \texttt{exoplanet} package and \texttt{PyMC3} \citep{exoplanet:pymc3} with the NUTS sampler \citep{Hoffman2011_NUTS}.
    The parameters describing the Keplerian orbit are $T_0$, $P$, $e$, and $\omega$.
    The transit model was described using the depth, $R_p^2 / R_\star^2$, duration, $T_{1,4}$, the impact parameter, $b$, and quadratic limb-darkening coefficients, ($u_\star$ and $v_\star$).
    For the RV modeling, we added $K$, which is the RV semi-amplitude. Following \citet{Fulton_RadVel_2018}, we adopted the reparametrization of $T_0, \ln P, \sqrt{e} \cos \omega,  \sqrt{e} \sin \omega$ to obtain $P>0$, $K>0$, while avoiding any overestimation of $e$. 
    We added a jitter $\sigma_\textrm{ESPRESSO}$ to the RV model, as well as an offset corresponding to the systemic velocity seen by ESPRESSO $\langle RV \rangle_\textrm{ESPRESSO}$, along with a linear trend parameter, $rvtrend$. 
    The prior distributions used for the model are given in Table \ref{tab:priors}. The prior on the stellar radius $R_\star$ comes from the analysis described in Sect. \ref{sec:stellar_params}. Priors for $R_p/R_\star$, $T_0$ and $\ln P$ come from the \texttt{Transit Least-Squares} analysis with wide priors. The remaining priors are uniform and wide. The quadratic limb-darkening coefficients were determined using the \texttt{LDTK} \citep{Parviainen_LDTK_2015}, while using the \citet{exoplanet:kipping13} parametrization for the TESS bandpass as $(u_{\star}, v_{\star}) = (0.4512, 0.1196) \pm (0.0005, 0.0015),$ and we fixed the priors to those values.

    We sampled the posterior distribution with an Markov chain Monte Carlo (MCMC) algorithm. 
    We set our run to have a burn-in of 1\,000 samples, 8\,000 samples, and 8 chains to explore the parameter space (Fig \ref{fig:cornerplot_post}). We checked for convergence using the improved $\hat{R}$ convergence diagnostic, described in \citet{R_hat2019}.
    
\subsection{Results}
\label{sec:results}
\begin{figure}[h]
    \resizebox{\hsize}{!}{\includegraphics{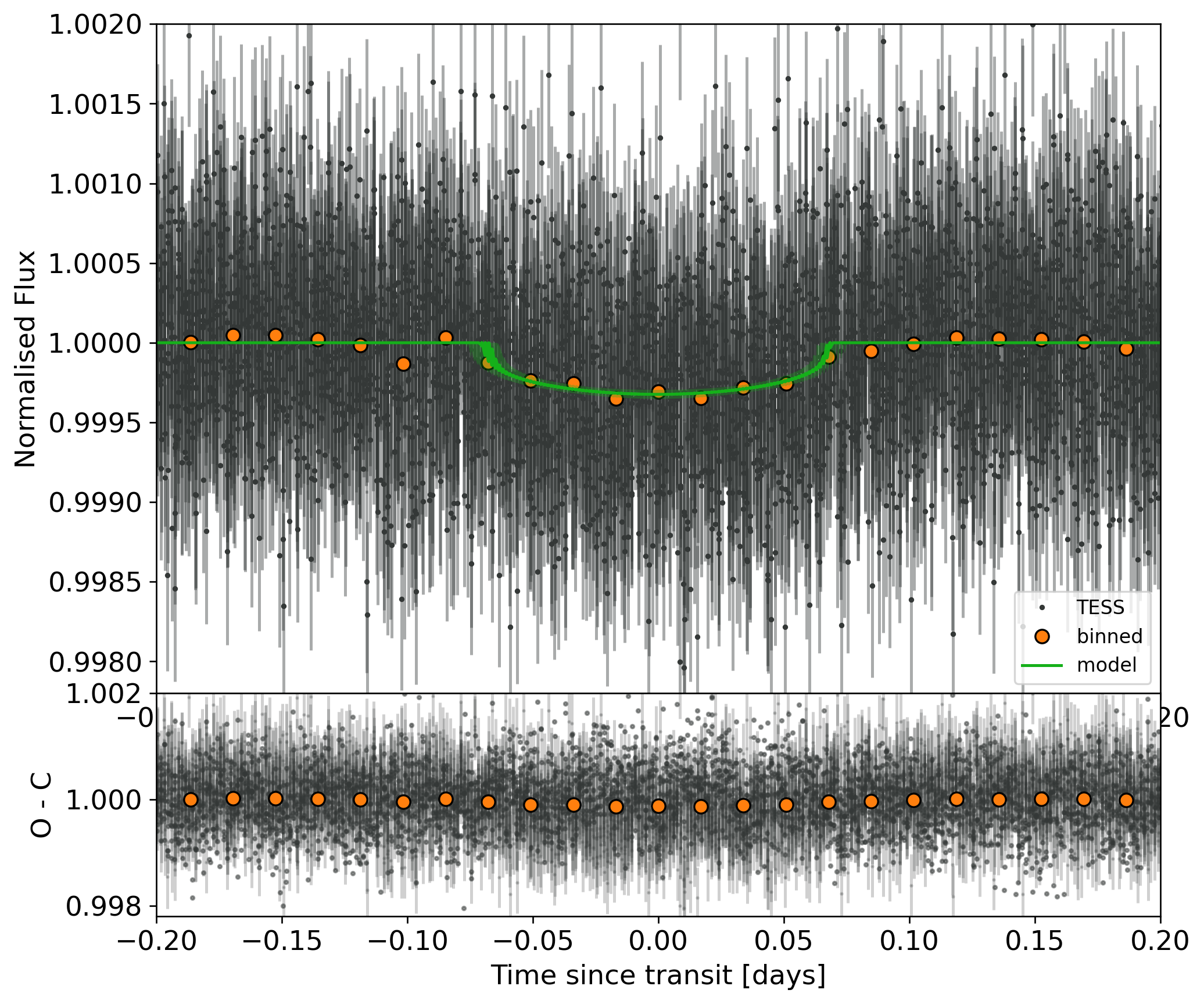}}
    \caption{Top: Phase-folded TESS light curve of TOI-512 at the period $P=7.18886$ days. Black dots represent TESS 2-min cadence data. For clarity, we plot the binned light curve in big orange dots. The solid green line and overlay correspond to our best-fit solution and its relative $1\sigma$ uncertainty.}
    \label{fig:lc_fold}
\end{figure}

    We evaluated the final parameters as the medians of the posterior distributions and the $1\sigma$ uncertainties as their $\pm~34.1\%$ quantiles. The results of our model sampling are listed in Table \ref{tab_toi512}. The phase-folded TESS light curve is shown on Fig. \ref{fig:lc_fold}.
    From our joint fitting of photometric and RV data, we find that TOI-512b is a super-Earth with a radius of $1.54 \pm 0.10$~R$_\oplus$ and mass of $3.54_{-0.54}^{+0.53}$~M$_\oplus$, on a 7.19~days orbit. This corresponds to a bulk density of~$5.28_{-1.19}^{+1.47}$~g~cm$^{-3}$. There is no evidence for non-zero eccentricity.

\begin{figure}[h!]
    \resizebox{\hsize}{!}{\includegraphics{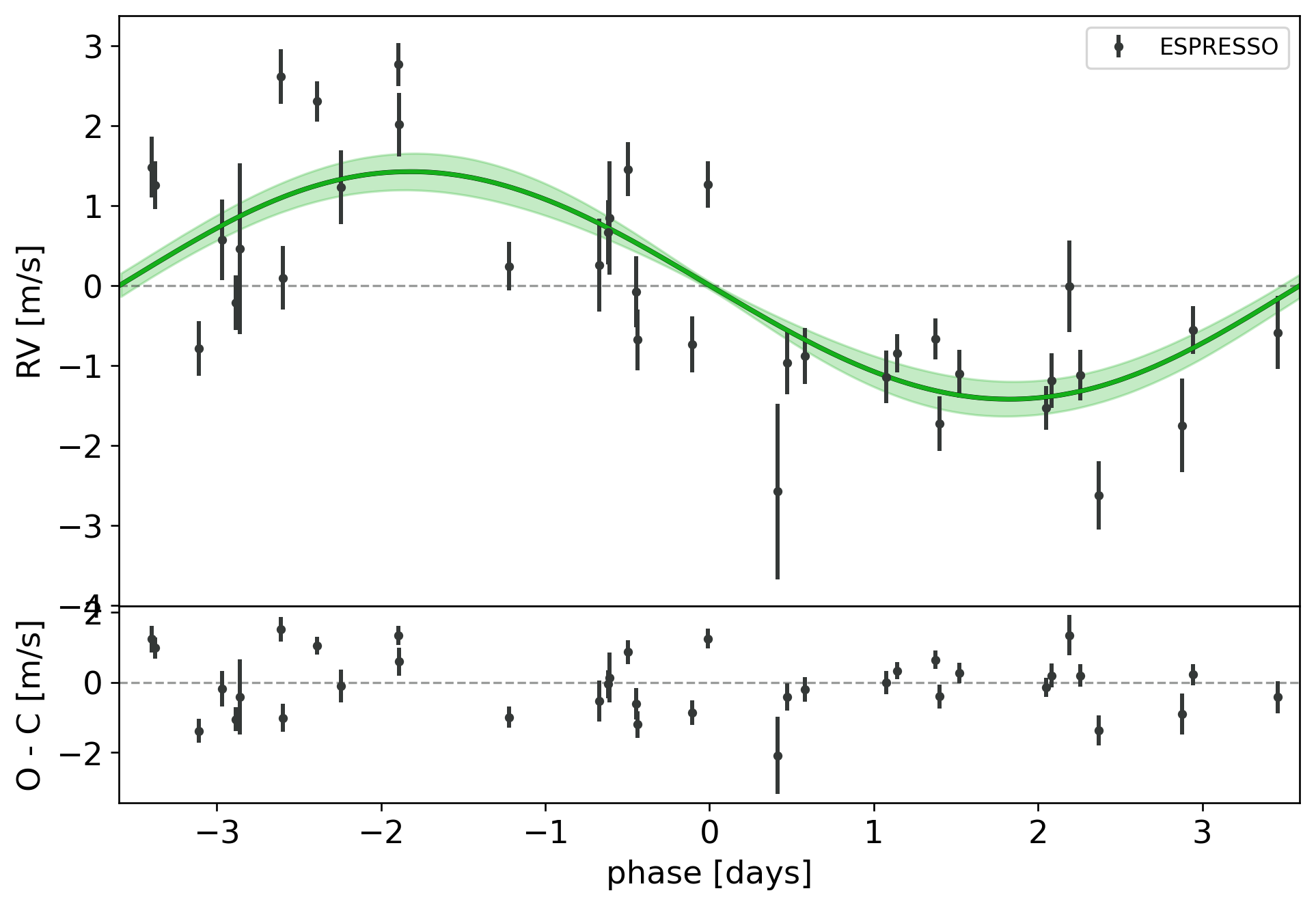}}
    \caption{Top: Phase-folded ESPRESSO observations of TOI-512 at the period $P=7.18886$~days. The solid green line and overlay correspond to our best-fit solution and its relative $1\sigma$ uncertainty. Bottom: The residuals of the best model.}
    \label{fig:rv_fold}
\end{figure}

\begin{table}
    \caption{Parameters of TOI-512b.}
    \begin{tabular}{l c}
    \hline
        Parameter & TOI-512b\\
        \hline
        Period (days) & $7.18886_{-7.7\cdot 10^{-5}}^{+6.9\cdot 10^{-5}}$\\\\

        $T_0$ (BJD-2 400 000) & $58471.212_{-0.004}^{+0.005}$\\\\

        Radius $R_p$ ($R_\oplus$) & $1.54\pm 0.10$ \\\\
        
        Mass $M_p$ ($M_\oplus$) & $3.57_{-0.55}^{+0.53}$ \\\\

        Density $\rho_p$ ($\rho_\oplus$) & $1.02_{-0.23}^{+0.29}$ \\\\
        
        $K$ (ms$^{-1}$) & $1.54 \pm 0.21$ \\\\
        
        $R_p / R_\star$ & $0.0157_{-9.6\cdot 10^{-4}}^{+9.8\cdot 10^{-4}}$ \\\\
       
        Impact parameter $b$ &{$0.30 \pm 0.2$ }\\\\

        Transit depth $\delta$ (ppm) & $246_{-30}^{+31}$ \\\\

        Transit duration $T_{1, 4}$ (h) &  $3.31_{-0.22}^{+0.20}$\\\\
        
        Eccentricity $e$ & $0.02_{-0.03}^{+0.06}$ \\\\

        Argument of periastron $\omega$ ($^\circ$) & $-54 \pm 150$\\\\
        
        Inclination $i$ ($^\circ$) & $88.92 \pm 0.66$ \\\\
        
        Insolation (Earth Flux) & $127_{-8}^{+9}$\\\\

        Equilibrium Temperature (K) & $1009 \pm 29$K\\
        
        \hline
        
    \end{tabular}
    \tablefoot{Equilibrium Temperature comes from the interior modeling described in Sec. \ref{sec:internal_structure}.}
    \label{tab_toi512}
\end{table}


\section{Mass-radius diagram and internal structure}
\label{sec:internal_structure}

\begin{figure}
        \resizebox{\hsize}{!}{\includegraphics{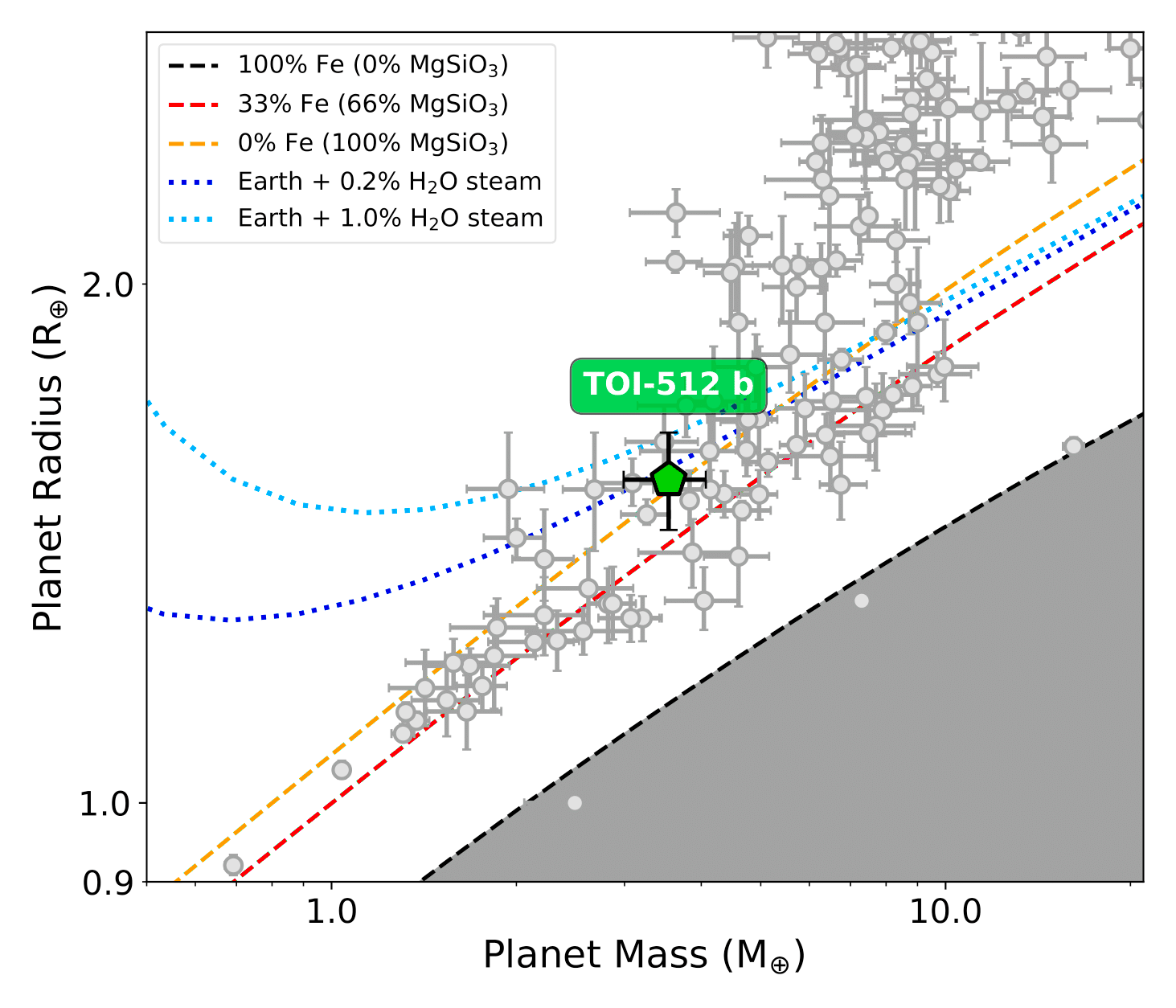}}
        \caption{Mass-radius diagram for planets with mass $<$30~$M_\oplus$ and radius $<$2.5~$R_\oplus$, selected from the \texttt{exoplanet.eu} archive \citep{Exoplanet_eu_Schneider_2011}. TOI-512b is the green pentagon. The overplotted composition models are from \citet{Zeng2016}. Steam atmosphere models are from \citet{Turbet2020}. The graph was created using the open-source code \texttt{mr-plotter} (\url{https://github.com/castro-gzlz/mr-plotter/}).}
        \label{fig:mass_radius_diag}
    \end{figure}

    Figure~\ref{fig:mass_radius_diag} shows the position of TOI-512b relative to the sample of planets from the \texttt{exoplanet.eu} archive \citep{Exoplanet_eu_Schneider_2011} with mass and radius measured with a relative precision lower than 20\% and 10\%, and relative to composition models obtained from \citet{Zeng2016} and \citet{Turbet2020}. TOI-512b stands near the MgSiO$_3$ line, consistent with a rocky composition and a thin steam atmosphere.

\begin{figure}
        \resizebox{\hsize}{!}{\includegraphics{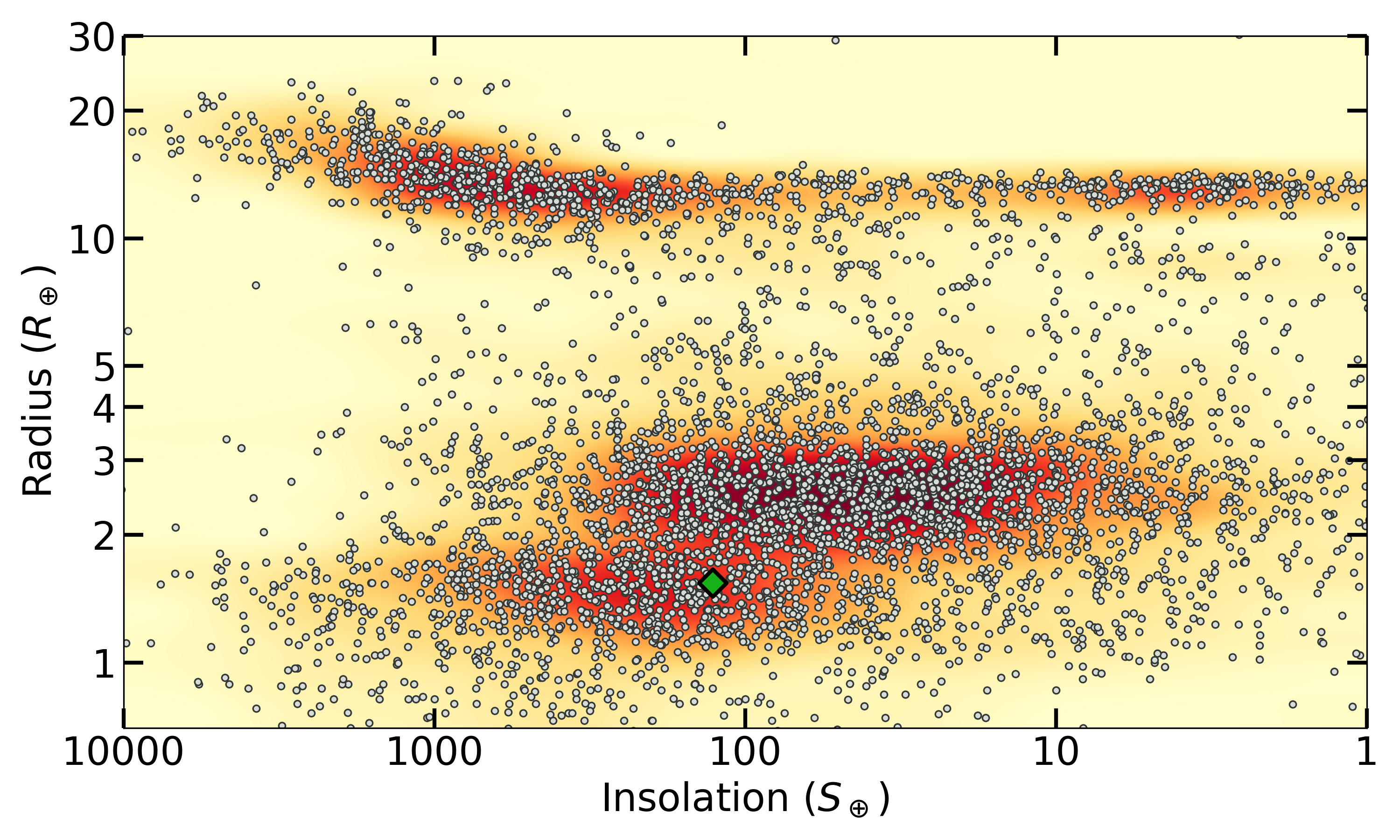}}
        \caption{Radius-insolation diagram for known planets, selected from the \texttt{exoplanet.eu} archive \citep{Exoplanet_eu_Schneider_2011}. Gray dots represent planets with known radius and insolation. TOI-512b is the green diamond.}
        \label{fig:radius_insolation}
    \end{figure}

    Furthermore, when visualized in a radius-insolation plot (Fig. \ref{fig:radius_insolation}), TOI-512b stands in the super-Earth overdensity, at an insolation level where both rocky and gaseous exoplanets exist. Given the age of the system ($8.235 \pm 4.386$~Gyr), such a result is expected and would be consistent with both XUV photoevaporation and core-powered mass loss. 
    
\begin{figure}
    \resizebox{\hsize}{!}{\includegraphics{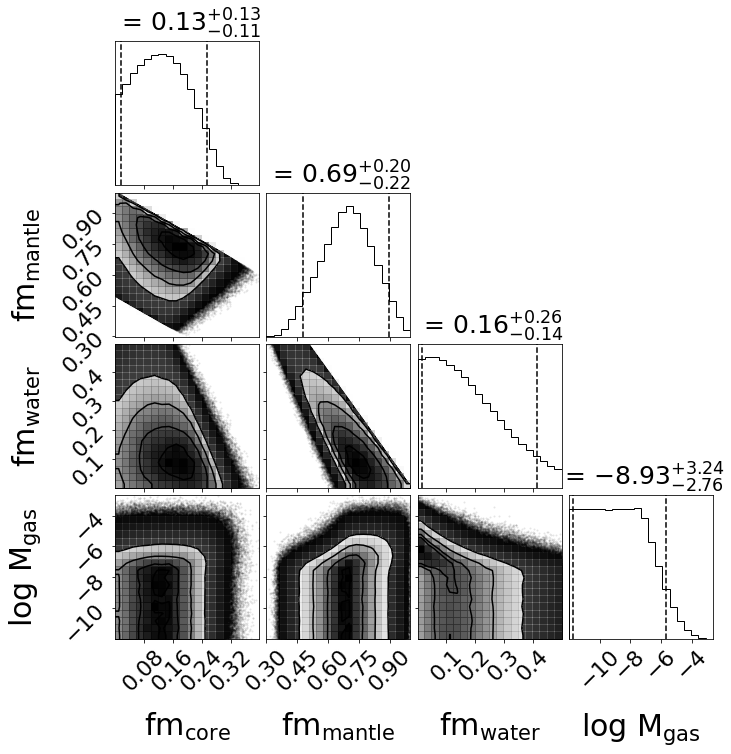}}
    \caption{Corner plot of the derived internal structure parameters of TOI-512b.}
    \label{fig:interior_corner}
\end{figure}

    We performed a Bayesian analysis to infer the internal structure of the planet TOI-512b, the same one already described in \citep{Dorn2015,Dorn2017} and used to study the internal structures of TOI-178 \citep{Leleu2021} or Nu$^2$ Lupi \citep{Delrez2021}.
    The model considers four layers: an inner iron core (Fe, S), a silicate mantle (Si, Mg and Fe), a water shell, and a gaseous envelope of pure H-He. The core, mantle, and water layer represent the solid part of the planet and are governed by the equations of state of \citep{Hakim2018}, \citep{Sotin2007}, and \citep{Haldemann2020} respectively.
    The Bayesian analysis was then carried out as follows: 5000 synthetic stars were generated, whose properties match those observed for the central star of the system in the error bars (i.e., mass, radius, effective temperature, age, and Mg/Si/Fe bulk molar ratios). Then, 2000 planets were produced around each star, whose properties  were allowed to vary, with free Mg/Si/Fe molar ratios. For each planet, the transit depth and RV semi-amplitude could then be calculated, and only the planets whose parameters fit the observables within their error bars were retained.
    The model simulates the mass fractions of each layer, the molar fraction of iron and sulfur in the core, silicon and magnesium in the mantle, the equilibrium temperature, and the age of the planet (equal to that of the star). Linearly uniform priors are taken for these parameters, except for the mass of the gas, which follows a uniform-in-log prior.
    The water fraction of the solid part is also estimated not to exceed 0.5. For a detailed explanation, we refer to \citep{Leleu2021}.
    The result of the modeling of the internal composition of the planet TOI-512b is presented in Fig.~\ref{fig:interior_corner}. Our Bayesian analysis results in a small inner core representing $0.13^{+0.13}_{-0.11}$ of the solid mass fraction for the planet, surrounded by a mantle with a mass fraction of $0.69^{+0.20}_{-0.22}$ and an upper limit of the water layer of $0.16$. A gas mass below $-8.93$ dex indicates a very small amount of gas on the planet.
    
    Forcing a model without water and without atmosphere failed to model a planet compatible with the data. Therefore, the planet certainly has water and/or an atmosphere. Due to its equilibrium temperature of $1009 \pm 29$K, we would expect a simple atmosphere made of H-He (like our model) to have been evaporated. We would also expect the water as a thick vapor atmosphere. Future spectroscopic observations would be required to solve this dichotomy.

    Our interior structure analysis is not compatible with the absence of water and an atmosphere, as would be expected from photoevaporation models for a planet this old. It is, however, consistent with atmospheric mass-loss driven by the core-powered mass loss mechanism, which assumes a maximum water mass fraction of $\sim 20\%$ and acts at the Gyr timescale \citep{GuptaSchlichting2019}.

\section{Spectroscopic follow-up observations}
\label{sec:spectroFollow-up}

    The brightness of the star makes TOI-512b a promising candidate for follow-up observations, although  the  small size of the planet does pose a challenge. To quantify how amenable our target would be  to atmospheric characterization, in particular with  James Webb Space Telescope \citep[JWST,][]{Gardner2023}, we employed the transmission spectroscopy metric (TSM) proposed by \citet{Kempton2018}. We made use of the scale factors provided in their Table 1 instead of the suggested value for temperate planets, since TOI-512~b does not fit into that category (insolation level $S_{\rm p} > 20~S_\oplus$). The planet radius of $R_{\rm p} = 1.54~\pm~0.10~R_\oplus$ overlaps within $1~\sigma$ the two smallest planet-size bins defined by \citet{Kempton2018}; namely, terrestrial planets and small sub-Neptunes, separated at $1.5~R_\oplus$. These two categories have substantially different scale factors and cutoffs for follow-up efforts, thus, we computed the TSM by sampling $R_{\rm p}$, $M_{\rm p}$, $T_{\rm eq}$, $R_\star$, and $m_J$ from their observational Gaussian distributions 1~000~000 times, then associating each individual throw with the corresponding scale factor. In this way, we were able to build two subsamples of TSM values: the first one $\rm{TSM}_{\rm Terr}$ linked to the terrestrial bin size and the second one $\rm{TSM}_{\rm Nept}$ to the small sub-Neptune bin. By computing the median and the 68\% highest density interval of the two subsamples, we obtained $\rm{TSM}_{\rm Terr} = 5^{+1}_{-1}$ and $\rm{TSM}_{\rm Nept} = 41^{+7}_{-10}$. Even though \citet{Kempton2018} recommended a cautious rectification of their metric for stars with $m_J < 9$, the two TSM values we get are significantly (at the $5~\sigma$ level) lower than the threshold they set for follow-up efforts, making our target a challenging case for atmospheric characterization observations with  JWST.

    It should be noted that thermal emission measurements at long wavelengths (i.e., $> 5$\,$\mu$m) with JWST/MIRI may be more advantageous than the transmission for the warmer terrestrial planets, as suggested by \citet{Morley2017}. Keeping the same procedure as before, we compute the emission spectroscopy metric (ESM), also proposed by \citet{Kempton2018}, by sampling $R_{\rm p}$, $T_{\rm eq}$, $T_{\rm eff}$, $R_\star$, and $m_K$ from their observational Gaussian distributions 1\,000\,000 0.152 times. Using the median and the 68\% HDI of the ensuing sample, we found a result of $\mathrm{ESM} = 2.8^{+0.4}_{-0.5}$, which is still far from the minimal requirement set by \citet{Kempton2018} for secondary eclipse spectroscopy, at the $\sim 12 \sigma$ level. This result further corroborates the potential difficulty for atmospheric characterization with  JWST.
    
    Additionally, we assessed the feasibility of studying the spin--orbit angle of TOI-512b via Rossiter-McLaughlin (RM) observations \citep{Rossiter1924,McLaughlin1924}, which gives access to important insights on the system's past dynamical evolution. In any case, we do not expect tidal realignment \citep[e.g.,][]{Barker2010} to be efficient, given the small size of the planet and its relatively long separation from the star. Indeed, we calculated the tidal efficiency factor to be $\sim 9 \times 10^{-19}$, following the prescription of \citet{Attia2023}; our result is below the threshold of $10^{-15}$ they set for considering efficient tidal realignment. On the other hand, the system is fairly old, which may still have given enough time for realignment to occur if the initial orbital tilt was mild enough. Another scenario would be a quiescent, coplanar, formation, and evolution, which could be corroborated by the seemingly circular nature of the orbit. 
    To this effect, we compute the expected maximum amplitude of the RM anomaly \citep[e.g.,][]{Winn2010} to be $\sim 0.38$ m/s. Using the classical, velocimetric, RM technique \citep[e.g.,][]{Ohta2005} would thus be extremely laborious to retrieve such a small signal. Newer methods, such as RM revolutions \citep{Bourrier2021}, might help analyze this challenging target. Providing constraints on the spin-orbit angle of TOI-512b would prove invaluable to disentangle these various past dynamical histories.


\section{Conclusions}
\label{sec:conclusion}
    In this paper we present the TOI-512 system and the discovery of the transiting super-Earth TOI-512b. We obtained 37 high-precision RV follow-up observations with the ESPRESSO spectrograph to measure the mass of TOI-512b and combined them with TESS photometry to estimate its composition. With a radius of $1.54 \pm 0.10$~R$_\oplus$ and a mass of $3.57_{-0.55}^{+0.53}$~M$_\oplus$, TOI-512b exhibits a rocky composition. However, our interior structure analysis is not compatible with the absence of water and an atmosphere, as would be expected from photoevaporation models for a planet this old. It is, however, consistent with atmospheric mass-loss driven by the core-powered mass loss mechanism, which assumes a maximum water mass fraction of $\sim 20\%$ and acts at the Gyr timescale \citep{GuptaSchlichting2019}.
    TOI-512b would be an interesting candidate for Rossiter-McLaughlin observations to further our understanding of planet formation and evolution. The transmission spectroscopy would be too tedious with the current generation of instruments, including the JWST. However, the quiet nature of the star will make it a good target for transmission spectroscopy with ANDES.
    We found no sign of the second planet proposed by TESS, neither in photometry, nor in radial velocities. A visual analysis of the light curve, phase-folded at the expected period, revealed oscillations that are not compatible with a transiting signal (Fig. \ref{fig:transitcheck}).


\begin{acknowledgements}
      The authors acknowledge the ESPRESSO project team for its effort and dedication in building the ESPRESSO instrument. 
      This research has made use of the NASA Exoplanet Archive, which is operated by the California Institute of Technology, under contract with the National Aeronautics and Space Administration under the Exoplanet Exploration Program. 
      This research has made use of data obtained from or tools provided by the portal exoplanet.eu of The Extrasolar Planets Encyclopedia.
      Funded/Co-funded by the European Union (ERC, FIERCE, 101052347). Views and opinions expressed are however those of the author(s) only and do not necessarily reflect those of the European Union or the European Research Council. Neither the European Union nor the granting authority can be held responsible for them.
      This work was financed by Portuguese funds through FCT (Funda\c c\~ao para a Ci\^encia e a Tecnologia) in the framework of the project 2022.04048.PTDC (Phi in the Sky, DOI 10.54499/2022.04048.PTDC). CJM also acknowledges FCT and POCH/FSE (EC) support through Investigador FCT Contract 2021.01214.CEECIND/CP1658/CT0001 (DOI 10.54499/2021.01214.CEECIND/CP1658/CT0001).
      This work was supported by FCT - Fundação para a Ciência - through national funds and by FEDER through COMPETE2020 - Programa Operacional Competitividade e Internacionalização by these grants: UID/FIS/04434/2019; UIDB/04434/2020; UIDP/04434/2020; PTDC/FIS-AST/32113/2017 \& POCI-01-0145-FEDER-032113; PTDC/FIS-AST/28953/2017 \& POCI-01-0145-FEDER-028953; PTDC/FIS-AST/28987/2017 \& POCI-01-0145-FEDER-028987;  EXPL/FIS-AST/0615/2021; PTDC/FIS-AST/30389/2017 \& POCI-01-0145-FEDER-030389.
      FPE and CLO would like to acknowledge the Swiss National Science Foundation (SNSF) for supporting research with ESPRESSO through the SNSF grants nr. 140649, 152721, 166227, 184618 and 215190. The ESPRESSO Instrument Project was partially funded through SNSF's FLARE Programme for large infrastructures.
      We thank the Swiss National Science Foundation (SNSF) and the Geneva University for their continuous support to our planet low-mass companion search programmes. 
      This work has been carried out within the framework of the NCCR PlanetS supported by the Swiss National Science Foundation under grants 51NF40\_182901 and 51NF40\_205606. J.D., M.A, M.H., F.B, and Y.A. acknowledge the financial support of the SNSF. M.A. has received funding from the European Research Council (ERC) under the European Union's Horizon 2020 research and innovation programme (project SPICE DUNE, grant agreement no. 947634; grmodeant agreement no. 730890).
      A.C.-G. is funded by the Spanish Ministry of Science through MCIN/AEI/10.13039/501100011033 grant PID2019-107061GB-C61.
      The INAF authors acknowledge financial support of the Italian Ministry of Education, University, and Research through PRIN 201278X4FL and the ``Progetti Premiali'' funding scheme.
      ASM, JIGH, RR, and CAP acknowledge financial support from the Spanish Ministry of Science, Innovation and Universities (MICIU) project PID2020-117493GB-I00.
      MRZO acknowledges financial support from project PID2022-137241NB-C42 approved by the Spanish Ministerio de Ciencia e Innovación.
      This research made use of \textsf{exoplanet} \citep{exoplanet:joss, exoplanet:zenodo} and its dependencies \citep{exoplanet:agol20, exoplanet:arviz, exoplanet:astropy13, exoplanet:astropy18, exoplanet:kipping13, starry2019, exoplanet:pymc3, exoplanet:theano, exoplanet:vaneylen19}.
      This research has made use of the Exoplanet Follow-up Observation Program (ExoFOP; DOI: 10.26134/ExoFOP5) website, which is operated by the California Institute of Technology, under contract with the National Aeronautics and Space Administration under the Exoplanet Exploration Program.
      We acknowledge financial support from the Agencia Estatal de Investigaci\'on of the Ministerio de Ciencia e Innovaci\'on MCIN/AEI/10.13039/501100011033 and the ERDF “A way of making Europe” through project PID2021-125627OB-C32, and from the Centre of Excellence “Severo Ochoa” award to the Instituto de Astrofisica de Canarias.
      This work has made use of data from the European Space Agency (ESA) mission {\it Gaia} (\url{https://www.cosmos.esa.int/gaia}), processed by the {\it Gaia} Data Processing and Analysis Consortium (DPAC, \url{https://www.cosmos.esa.int/web/gaia/dpac/consortium}). Funding for the DPAC has been provided by national institutions, in particular the institutions participating in the {\it Gaia} Multilateral Agreement.

\end{acknowledgements}

\bibliographystyle{aa}
\bibliography{main}


\begin{appendix} 

\section{Prior distributions}
\begin{table}[h!]
\caption{Prior distributions of Sect. \ref{sec:joint_analysis}}
\begin{tabular}{l c c}
    \hline
    Parameter & Prior & Value\\
    \hline
    Stellar parameters: &  & \\
    $R_\star$ (R$_\odot$) & $\mathcal{N}(0.9, 0.03)$ & $0.89 \pm 0.03$\\
    $M_\star$ (M$_\odot$) & $\mathcal{N}(0.74, 0.1)$ & $0.74 \pm 0.03$\\
    ($u_\star$, $v_\star$)  & \citet{exoplanet:kipping13}  & $(0.4512, 0.1196) \pm (0.0005, 0.0015)$\\
    $\langle RV \rangle_\textrm{ESPRESSO}$ $(m  s^{-1})$ & $\mathcal{N}(68076, 10)$ & $68076.96 \pm 0.14$\\
    $rvtrend$ $(m  s^{-1} yr^{-1})$ & $\mathcal{U}(-5,5)$ & $1.24 \pm 0.34$\\\\

    Transit Parameters: &  & \\
    $R_p / R_\star$ & $\mathcal{U}(0.001, 0.1, 0.01)$  & $0.0157_{-9.6\cdot 10^{-4}}^{+9.8\cdot 10^{-4}}$ \\
    $b$ & $\mathcal{U}(0, 1, 0.2)$  & $0.34 \pm 0.21$\\\\

    Keplerian parameters: &  & \\
    $T_0$ (BJD-2 400 000) & $\mathcal{N}(58471.2, 1)$ & $58471.212_{-0.004}^{+0.005}$\\
    $\ln P$ & $\mathcal{N}(1.97, 2)$ &  $1.973^{+7.93\cdot10^{-6}}_{-10^{-5}}$\\
    $\ln K$ & $\mathcal{U}(-1.6, 1.6)$ & $0.432^{+0.139}_{-0.161}$\\
    $\sqrt{e}\cos \omega$ & $\mathcal{U}(-1,1)$ & $-0.001_{-0.398}^{+0.395}$\\
    $\sqrt{e}\sin \omega$ & $\mathcal{U}(-1,1)$ & $0.006_{-0.389}^{+0.386}$\\
    
    \hline
    
\end{tabular}

\tablefoot{Numbers in brackets represent: \\
(lower limit $x$, upper limit $y$, test value $z$) for uniform distributions $\mathcal{U}(x, y, z)$. \\
(mean $\mu$, standard deviation $\sigma$, test value $\alpha$) for normal distributions $\mathcal{N}(\mu, \sigma, \alpha)$.}

\label{tab:priors}
\end{table}

\vspace*{-5ex}

\section{FULMAR}

With the advent of all sky photometric surveys looking for exoplanets, a large number of interesting exoplanet transit candidates have been detected in a short amount of time.
Given the high instrumental time required for an efficient radial velocity follow-up, most of these targets cannot realistically be observed using ground facilities. Our tool is aimed at helping astronomers select suitable RV follow-up targets more effectively and making their analysis easier and more convenient. By combining information coming from photometric observations with spectroscopic ones, one might reduce the necessary observation time per target, allowing for the characterization of more systems with a given instrument. Current publicly available candidates can benefit from a more detailed analysis to whether confirm the candidates, check for false positives, and find other transits in the system. A careful correction of the stellar activity can be necessary for the latter, and to help with the follow-up strategy. We also aimed for practicality, by using a single tool to retrieve the light curves and stellar parameters, correct the activity, look for exoplanets, and estimate their fundamental parameters.

FULMAR is an open source Python package that was created to assist astronomers involved in radial velocity follow-up programs of transiting exoplanets candidates coming from space surveys, such as \textit{Kepler} and TESS, by making the analysis of the light curves easier. It provides tools to download light curves, correct stellar activity in several ways, to look for transits, to refine transit parameters, to estimate the amplitude of the corresponding RV signal, and to visually probe signals detected in RV. It was build in a modular way, making new features easier to implement. FULMAR is currently used in the working group 3 ("RV follow-up of K2 and TESS transiting planets") of the ESPRESSO consortium \citep{Pepe_espresso_vlt_2021} for the selection and the preliminary analysis of candidates.\\
\vfill\null

\vspace*{10.32cm}

\section{Extra plot}

\begin{figure}[h!]
    \includegraphics[width=\hsize]{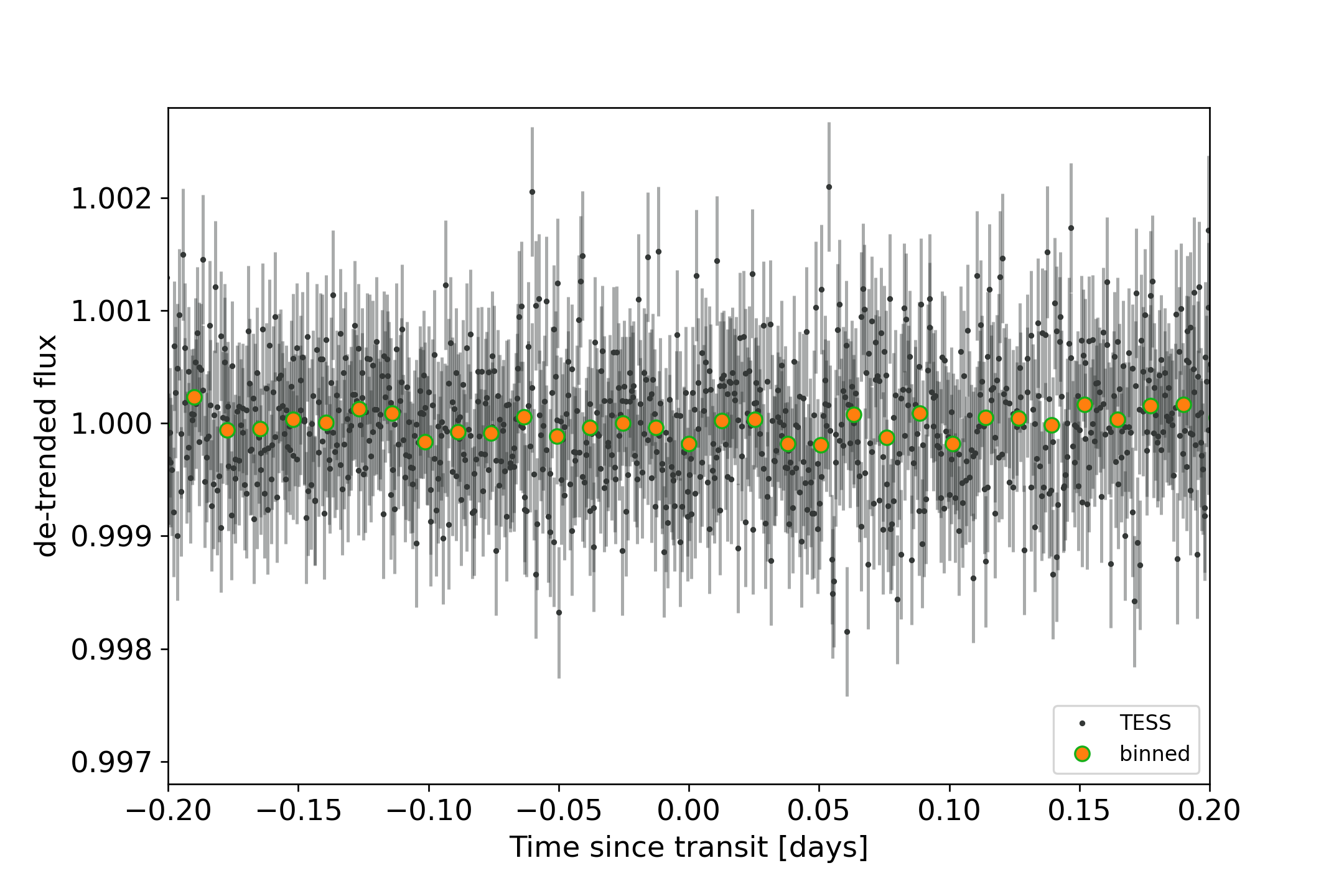}
    \caption{Phase-folded light curve at the expected period for TOI-512.02.}
    \label{fig:transitcheck}
\end{figure}    

\onecolumn
\section{Joint modeling of the posterior distributions}

\begin{figure*}[h!]
    \centering
    \includegraphics[width=\textwidth]{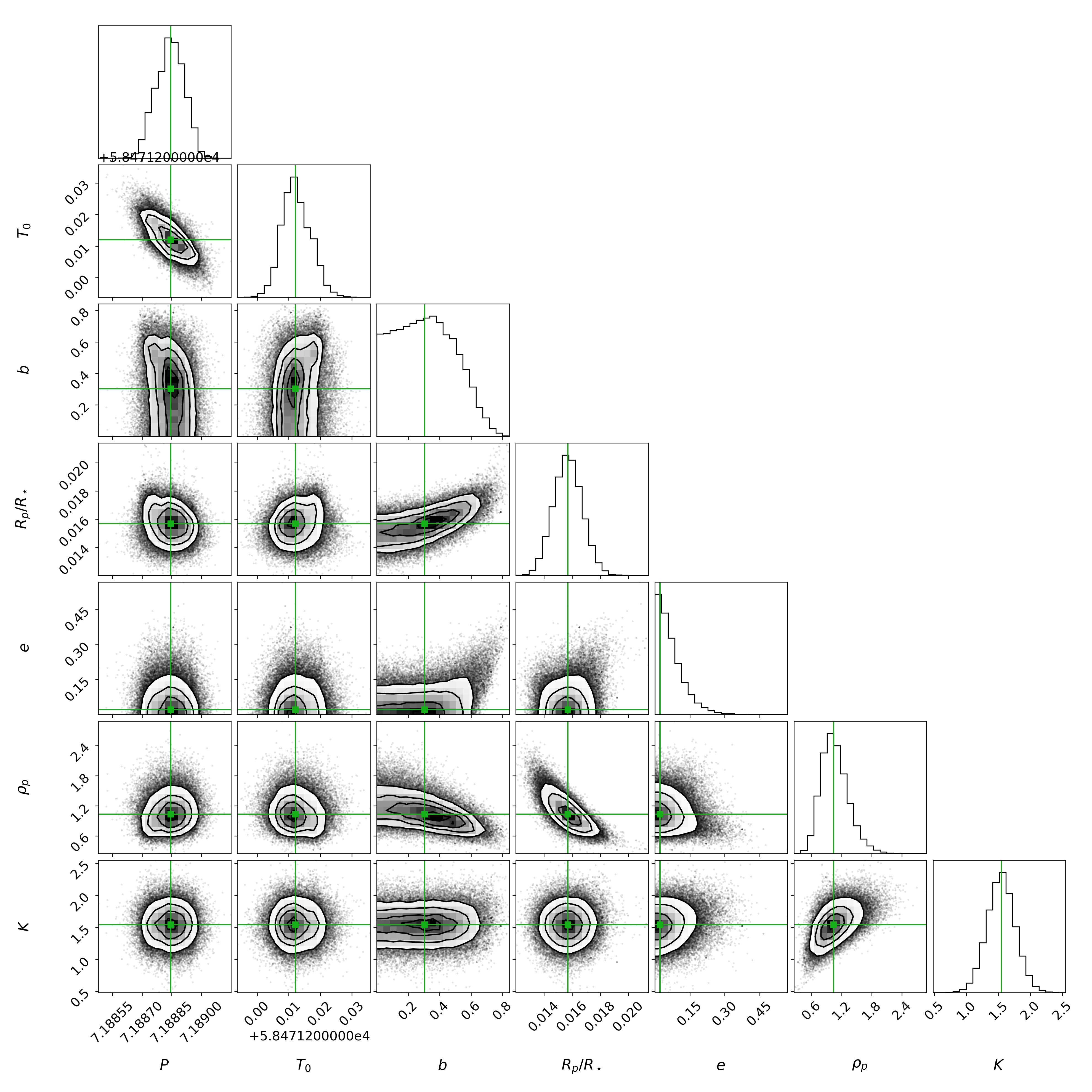}
    \caption{Posterior distribution for fitted parameters from the joint fit of TOI-512 b}
    \label{fig:cornerplot_post}
\end{figure*}

\end{appendix}

\end{document}